\documentclass[11pt,twoside]{article}
\usepackage{asp2010}

\resetcounters

\begin{document}

\title{X-ray Evidence for Ultra-fast Outflows in Local AGNs}
\author{F. Tombesi$^{1,2}$, M. Cappi$^3$, R.~M. Sambruna$^4$, J.~N. Reeves$^5$, C.~S. Reynolds$^2$, V. Braito$^6$, and M. Dadina$^3$
\affil{$^1$X-ray Astrophysics Laboratory and CRESST, NASA/GSFC, Greenbelt, MD 20771, USA}
\affil{$^2$Dept. of Astronomy, University of Maryland, College Park, MD 20742, USA}
\affil{$^3$INAF-IASF Bologna, Via Gobetti 101, I-40129 Bologna, Italy}
\affil{$^4$Dept. of Physics and Astronomy, 4400 University Drive, George Mason University, Fairfax, VA 22030}
\affil{$^5$Astrophysics Group, School of Physical and Geographical Sciences, Keele University, Keele, Staffordshire ST5 5BG, UK}
\affil{$^6$INAF-Osservatorio Astronomico di Brera, via. E. Bianchi 46, Merate, Italy}}

\begin{abstract}
X-ray evidence for ultra-fast outflows (UFOs) has been recently reported in a number of local AGNs through the detection of blue-shifted Fe XXV/XXVI absorption lines. We present the results of a comprehensive spectral analysis of a large sample of 42 local Seyferts and 5 Broad-Line Radio Galaxies (BLRGs) observed with \emph{XMM-Newton} and \emph{Suzaku}. We detect UFOs in $\ga$40\% of the sources. Their outflow velocities are in the range $\sim$0.03--0.3c, with a mean value of $\sim$0.14c. The ionization is high, in the range log$\xi$$\sim$3--6~erg~s$^{-1}$~cm, and also the associated column densities are large, in the interval $\sim$$10^{22}$--$10^{24}$~cm$^{-2}$. Overall, these results point to the presence of highly ionized and massive outflowing material in the innermost regions of AGNs. Their variability and location on sub-pc scales favor a direct association with accretion disk winds/outflows. This also suggests that UFOs may potentially play a significant role in the AGN cosmological feedback besides jets and their study can provide important clues on the connection between accretion disks, winds and jets.
\end{abstract}

\section{Introduction}

The recent detection of blue-shifted Fe~XXV/XXVI absorption lines in the X-ray spectra of several Seyferts and quasars suggests the presence of highly ionized and mildly-relativistic outflows in the center of these AGNs (e.g., Chartas et al.~2002, 2003; Pounds et al.~2003; Dadina et al.~2005; Markowitz et al.~2006; Braito et al.~2007; Cappi et al.~2009; Reeves et al.~2009; Giustini et al.~2011). They are possibly directly connected with accretion disk winds/outflows and the high associated outflow rate and mechanical power suggest they might be able to provide an important contribution to the expected AGN cosmological feedback (e.g., King 2010; Tombesi et al.~2012). Here we describe the systematic analysis and characterization of these so called Ultra-fast Outflows (UFOs) on a large sample of both radio-quiet and radio-loud AGNs. These results are discussed in detail in Tombesi et al.~(2010a, b; 2011a, b).

    \begin{figure}[!t]
   \centering
    \includegraphics[width=5.5cm,height=5cm,angle=0]{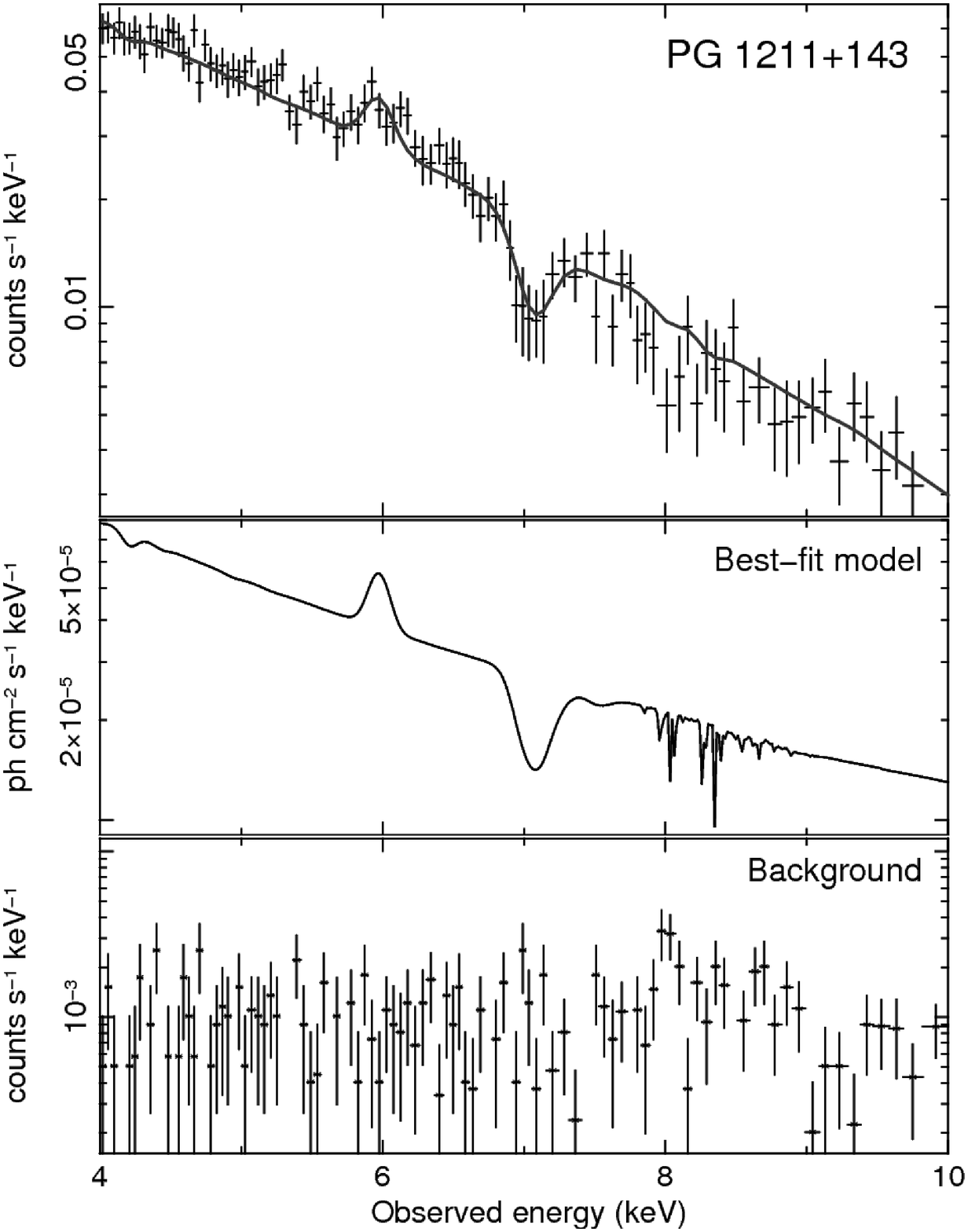}
\hspace{0.2cm}
    \includegraphics[width=5.5cm,height=4cm,angle=0]{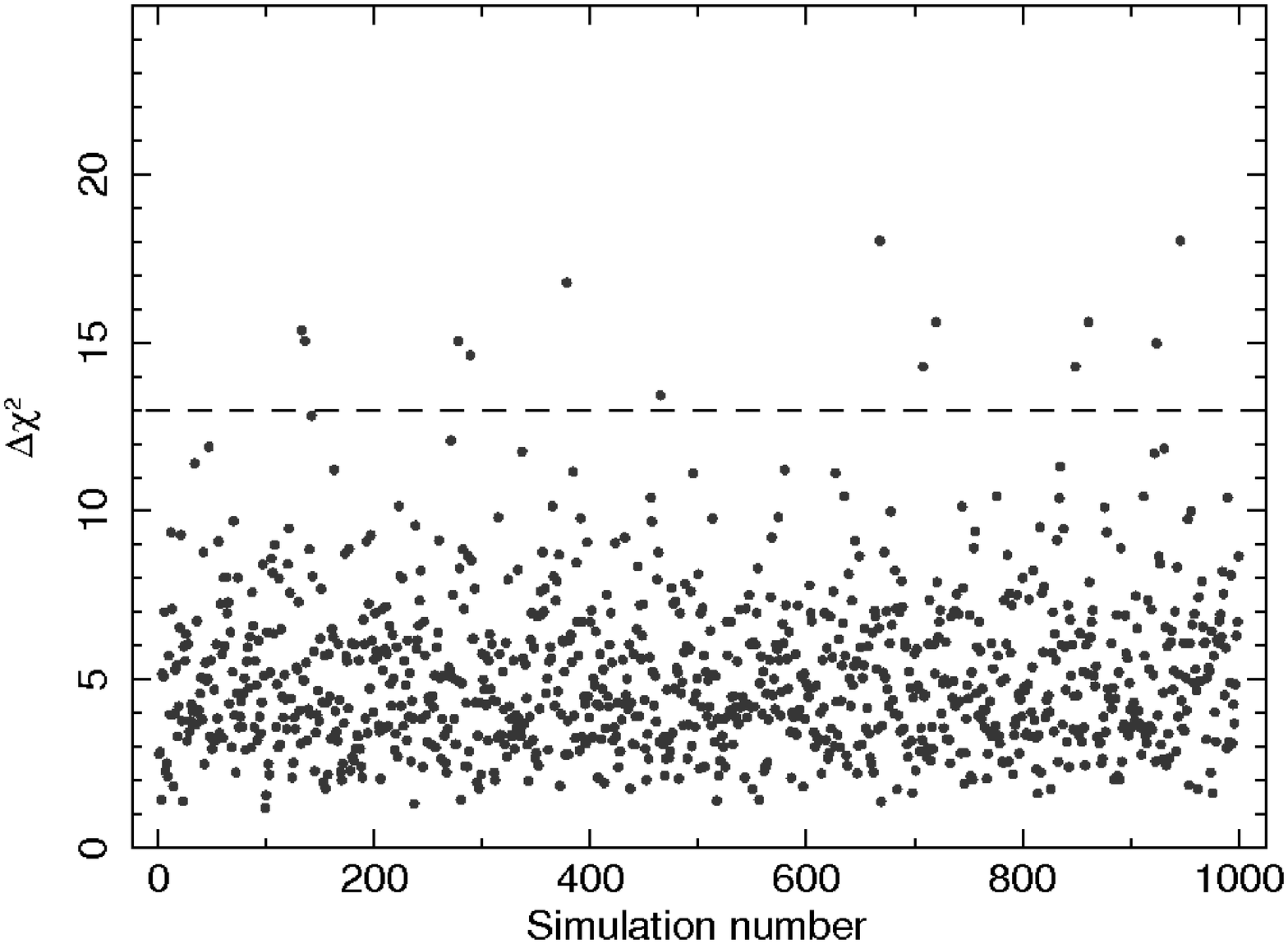}
   \caption{\emph{Left panel:} XMM-Newton EPIC-pn spectrum of PG~1211$+$143, best-fit model and background spectrum. \emph{Right panel:} Example of $\Delta\chi^2$ distribution from 1000 Monte Carlo simulations.}
    \end{figure}

\section{The Radio-Quiet Sample}

The radio-quiet sample was defined in Tombesi et al.~(2010a) selecting all the Narrow-Line Seyfert 1, Seyfert 1 and Seyfert 2 (with neutral absorption column density $N_H<10^{24}$ cm$^{-2}$) in the RXTE All-Sky Slew Survey Catalog (Revnivtsev et al.~2004) and then cross-correlated with the \emph{XMM-Newton} catalog. We obtained 42 sources for a total of 101 pointed \emph{XMM-Newton} observations. The sources are local ($z$$\le$0.1) and X-ray bright. We applied standard screening procedures and extracted the 4--10~keV EPIC pn spectra. We carried out a uniform spectral analysis using a phenomenological baseline model composed of an absorbed power-law continuum and Gaussian Fe K emission lines (see left panel of Fig.~1). Then, we performed a blind rearch for emission/absorption lines adding an additional narrow line to the baseline models with free positive/negative intensity and stepped the line energy between 4--10~keV, recording each time the associated $\Delta\chi^2$ deviations. The resultant energy-intensity contour plots were inspected to select possible features. Initially, only the narrow lines ($\sigma$$\le$100~eV) with $>$99\% F-test detection probability were selected and their velocity shifts were estimated using an identification as Fe~XXV/XXVI 1s--2p/1s--3p transitions. As discussed in \S4, we subsequently performed additional tests on the significance of the absorption lines at E$\ge$7.1~keV using extensive Monte Carlo simulations and selected only those with probability $\ge$95\% (see right panel of Fig.~1). Thus, in Tombesi et al.~(2010a) we detected a total of 36 absorption lines, 14 in the range E$=$6.4--7.1~keV and 22 at E$\ge$7.1~keV.

In a subsequent paper, Tombesi et al.~(2011a), we then performed a detailed curve of growth analysis of the Fe~XXV/XXVI absorption lines and a photo-ionization modeling using the \emph{Xstar} code (see left panel of Fig.~1). We considered an ionizing continuum equivalent to the average SED of the Seyfert 1s in the sample, i.e. a $\Gamma$$\simeq$2 power-law with cut-off at E$\simeq$100~keV. We then calculated \emph{Xstar} grids assuming standard Solar abundances and turbulent velocities of 1000, 3000 and 5000~km/s. Given the limited energy resolution of the EPIC pn, only four absorption lines were resolved with width $\sigma$$\simeq$5,000~km/s and for the other we could place only upper limits. 
A blind search for the best-fit \emph{Xstar} solution(s) was performed stepping the absorber redshift between 0.1 and $-$0.4, leaving the column density and ionization free, and checking for minima in the resultant $\chi^2$ distribution. This allows to self-consistently take into account the possible presence of lines/edges from ions of different elements. In all cases a good fit (P$_F$$\ga$99\%) was reached with a single \emph{Xstar} component and all the absorption lines were indeed consistent with blue-shifted Fe~XXV/XXVI transitions. The possible degeneracy of the identification in a few cases was included in the relative larger parameter errors. 

These studies allowed us to derive the distribituion of the main parameters of the UFOs in the radio-quiet sample (Tombesi et al.~2011a). The detection fraction is $\ga$40\%, which may suggest a large covering factor of $\ga$0.4. The lines have been found to be variable in both EW and velocity shift on time-scales even as short as $\sim$days, indicating compact absorbers. The outflow velocities are mildly-relativistic, in the range $\sim$0.03--0.3c, with a mean value of $\sim$0.14c. Their ionization is high, in the interval log$\xi$$\sim$2.5--6~erg~s$^{-1}$~cm, and the associated column densities are also large, in the range $N_H$$\sim$$10^{22}$--$10^{24}$~cm$^{-2}$. The estimated location and energetics of the UFOs in the radio-quiet sample is reported in a subsequent paper, Tombesi et al.~(2012).

\section{The Radio-Loud Sample}

We extented the search for UFOs also in radio-loud AGNs and in particular to BLRGs, which are the radio-loud counterparts of Seyfert 1s. They show strong relativistic radio jets, but the typical inclination of $i$$\sim$20$^{\circ}$--40$^{\circ}$ allows the direct observation of the inner disk in X-rays. Therefore, in Tombesi et al.~(2010b) we performed a systematic 4--10~keV spectral analysis of the long \emph{Suzaku} observations of 3C~111, 3C~390.3, 3C~120, 3C~382 and 3C~445. We applied the same method as for the radio-quiet sample (see \S2) and detected blue-shifted absorption lines at E$>$7~keV in 3/5 sources. Their significance was assessed using both F-test and extensive Monte Carlo simulations (see \S4). Through a photo-ionization modeling with \emph{Xstar} we find that they are consistent with series of Fe~XXV/XXVI K-stell transitions blue-shifted with mildly-relativistic velocities in the range $\sim$0.04--0.15c. They are highly ionized, log$\xi$$\simeq$4--6~erg~s$^{-1}$~cm, and the associated column densities are $N_H$$\ga$$10^{22}$~cm$^{-2}$. Their estimated location within $\sim$0.01--0.1~pc from the central super-massive black hole suggests a likely origin related with accretion disk outflows. The mass outflow rate of these UFOs is comparable to the accretion rate of $\sim$1~$M_{\odot}$~yr$^{-1}$. Their estimated kinetic power is high, in the range $\sim$$10^{43}$--$10^{45}$~erg~s$^{-1}$, which is comparable to their typical jet power and corresponds to a significant fraction of the bolometric luminosity.  Therefore, these UFOs can potentially play a significant role in the expected AGN feedback and can also possibly be directly linked to the jet activity.

The galaxy 3C~111 was then the subject of a follow-up study of the variability of its UFO with \emph{Suzaku} in September 2010 (Tombesi et al.~2011b). We obtained three pointings of $\sim$60~ks spaced by $\sim$7~days and observed a $\sim$30\% flux increase between the first and the second. A Galactic absorbed power-law continuum plus a narrow Fe K$\alpha$ emission line at E$\simeq$6.4~keV provide a good representation of the 4--10~keV XIS spectra. However, an additional emission line at E$\simeq$6.88~keV is observed in the first observation and an absorption line at E$\simeq$7.75~keV in the second. The detection significance of these features is high, $>$99\% from both F-test and Monte Carlo simulations. They are also significantly variable among the observations.
 The emission feature can be modeled with a highly ionized line coming from reflection off the accretion disk at $\sim$20--100$r_g$ from the black hole. Instead, a photo-ionization modeling of the absorption line suggests its association with a highly ionized, log$\xi$$\simeq$4.3~erg~s$^{-1}$~cm, UFO with outflow velocity $\sim$0.1c and column density $N_H$$\simeq$$8\times 10^{22}$~cm$^{-2}$. The location of the material is constrained at $<$$0.006$~pc from variability.
This observation may provide the first direct evidence for an accretion disk-wind connection in an AGN and is consistent with a picture in which an increased illumination of the inner disk causes an outflow to be lifted at $\sim$100~$r_g$. This is then possibly accelerated through radiation pressure to the observed terminal velocity of $\sim$0.1c, but additional magnetic thrust can not be excluded (e.g., Ram{\'{\i}}rez \& Tombesi 2012).

   \begin{figure}[!t]
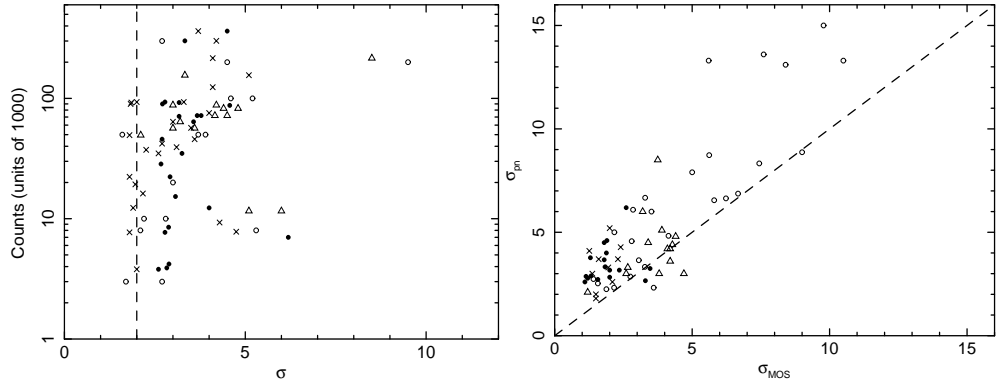

   \centering
    \includegraphics[width=5cm,height=6.5cm,angle=-90]{tombesi_fig3.ps}
%\hspace{0.2cm}
    \includegraphics[width=5cm,height=6.5cm,angle=-90]{tombesi_fig4.ps}
   \caption{\emph{Left panel:} significance of the Fe K lines detected by Tombesi et al.~(2010a) with respect to the 4--10~keV counts of the relative XMM-Newton EPIC pn observations. \emph{Right panel:} Comparison between the significance of the lines simultaneously detected in the EPIC pn and MOS (see \S4 for more details).}
    \end{figure}

\section{Significance of the blue-shifted absorption lines}

The significance of the absorption lines detected at E$\ge$7.1~keV was further investigated through extensive Monte Carlo simulations (Tombesi et al.~2010a, b). For each case we simulated 1000 spectra assuming the baseline model without the absorption lines and calculated the $\Delta\chi^2$ distribution of random generated features in the 7.1--10~keV band. Then, we selected only the observed absorption lines with measured $\Delta\chi^2$ corresponding to a Monte Carlo detection probability $\ge$95\%. We can also exclude any contamination due to the EPIC pn background and calibration uncertainties and we checked that the possible presence of spectral complexities, such as reflection and warm absorption, are typically weak enough to assure a only marginal model dependency of the results.   
This systematic analysis on a large sample of sources allows to estimate the global detection probability of the absorption lines of $>$5$\sigma$ and to overcame the possible publication bias claimed by Vaughan \& Uttley (2008). Moreover, we checked that the results are consistent using also the MOS detectors and the global random probability in that case is also low, $<$$10^{-7}$.

In the left panel of Fig.~2 we show the distribution of the significance $\sigma$ (simply estimated from the ratio $<$EW$>$/error$<$EW$>$) of the blue-shifted absorption lines at E$\ge$7.1~keV detected by Tombesi et al.~(2010a) with respect to the 4--10~keV counts of the relative observations (filled circles).
It has been claimed that the tendency of the distribution to lay close to the 3$\sigma$ level and not showing a strong enhancement in the significance following an increase in counts might suggest that all the detections are fake.
However, plotting also the same distribution for the absorption lines at E$=$6.4--7.1~keV (open triangles) and the ionized Fe K emission lines at E$=$6.5--7~keV (crosses), we note that they also follow this general trend and it is not peculiar of the E$\ge$7.1~keV lines.
In particular, we state that this does not necessarily mean that the lines are fake, given that their significance was already estimated using the F-test and extensive Monte Carlo simulation, but that they are intrinsically weak and variable.  
We support this statement with a simple test. We simulated 15 absorption lines assuming the same parameters as the real data, i.e. 4--10~keV counts in the range 3--300$\times 10^3$ and variable EW$=$20, 40, 60~eV and E$=$6. 8, 9~keV. As expected, the same trend is followed also by the simulations (open circles). 
Instead, in the right panel of Fig.~2 we show the distribution of the significance of the lines detected simultaneously in the EPIC pn and MOS. In this case the open circles refer to the narrow 6.4~keV Fe K$\alpha$ emission lines.
We can see that the points are systematically indicating that the detection significance in the pn is higher than in the MOS. This trend is simply expected from the higher effective area and lower background of the EPIC pn with respect to the MOS.  

Important improvements in the detection and characterization of these lines are expected from the higher effective area and supreme energy resolution in the Fe K band offered by the micro-calorimeters on board \emph{Astro-H} and the proposed ESA \emph{Athena} missions.

\acknowledgements FT thank G.~G.~C. Palumbo and R.~F. Mushotzky for the useful discussions.

\end{document}